\documentclass[%
 reprint,
showkeys,
 amsmath,amssymb,
 aps,
prd,
]{revtex4-1}

\usepackage{graphicx}
\usepackage{xcolor}
\usepackage{dcolumn}
\usepackage{bm}
\usepackage{slashed,float}

\begin{document}


\title{Thermal corrections to the gluon magnetic Debye mass}

\author{Alejandro Ayala$^{1,2}$,
Jorge David Casta\~no-Yepes$^1$,
        C. A. Dominguez$^2$,\\ 
        S. Hern\'andez-Ortiz$^1$, 
        L. A. Hern\'andez$^{1,2}$, 
        M. Loewe$^{2,3,4}$,
        D. Manreza Paret$^5$ and 
        R. Zamora$^{6,7}$}
 \affiliation{$^1$Instituto de Ciencias Nucleares, Universidad Nacional Aut\'onoma de M\'exico, Apartado Postal 70-543, CdMx 04510, Mexico.\\
 $^2$Centre for Theoretical and Mathematical Physics, and Department of Physics, University of Cape Town, Rondebosch 7700, South Africa.\\
 $^3$Instituto de F\'isica, Pontificia Universidad Cat\'olica de Chile, Casilla 306, Santiago 22, Chile.\\
 $^4$Centro Cient\'ifico-Tecnol\'ogico de Valpara\'iso CCTVAL, Universidad T\'ecnica Federico Santa Mar\'ia, Casilla 110-V, Valapara\'iso, Chile. \\
 $^5$Facultad de F\'isica, Universidad de La Habana, San Lazaro y L, La Habana, Cuba.\\
 $^6$Instituto de Ciencias B\'asicas, Universidad Diego Portales, Casilla 298-V, Santiago, Chile.\\
 $^7$Centro de Investigaci\'on y Desarrollo en Ciencias Aeroespaciales (CIDCA), Fuerza A\'erea de Chile,  Santiago, Chile.}


\begin{abstract}
We compute the gluon polarization tensor in a thermo-magnetic environment in the
strong magnetic field limit at zero and high temperature. The magnetic field effects are introduced using Schwinger's proper time method. Thermal effects are computed in the HTL approximation. At zero temperature, we reproduce the well-known result whereby for a non-vanishing quark mass, the polarization tensor reduces to the parallel structure and its coefficient develops an imaginary part corresponding to the threshold for quark-antiquark pair production. This coefficient is infrared finite and simplifies considerably when the quark mass vanishes. Keeping always the field strength as the largest energy scale, in the high temperature regime we analyze two complementary hierarchies of scales: $q^2\ll m_f^2\ll T^2$ and $m_f^2\ll q^2\ll T^2$.  In the latter, we show that the polarization tensor is infrared finite as $m_f$ goes to zero. In the former, we discuss the thermal corrections to the magnetic Debye mass.
\end{abstract}

\keywords{gluon polarization tensor, magnetic fields, finite temperature}
\maketitle


\section{\label{sec:level1} Introduction}

$\mathfrak{G}$
The properties of strongly interacting matter immersed in a magnetized medium have been the subject of intense research over the last years. The motivation for this activity stems from several fronts: On the one hand, lattice QCD (LQCD)~\cite{LQCD} has shown that for temperatures above the chiral restoration pseudo-critical temperature, the quark-antiquark condensate decreases and that this temperature itself also decreases, both as functions of the field intensity. This result, dubbed inverse magnetic catalysis (IMC), has sparked a large number of explanations~\cite{Bruckmann,Farias,Ferreira,Ayala0, Ayala1,Ayala2,Ayala3,Avancini,Ayala4,vertex1,vertex2,Mueller}. On the other hand, it has been argued that intense magnetic fields can be produced in peripheral heavy-ion collisions. Possible signatures of the presence of such fields in the interaction region can be the chiral magnetic effect~\cite{CME} or the enhanced production of prompt photons~\cite{photons, Skokov,Zakharov,Tuchin}. Moreover, magnetic fields can have an impact on the properties of compact astrophysical objects, such as neutron stars~\cite{Daryel}. 

The dispersive properties for gluons propagating in a magnetized medium are encoded in the gluon polarization tensor. For QED, this tensor has been computed and extensively studied both at zero and finite temperature~\cite{Dittrich,Hattori1,Hattori2,Fukushima,Ferrer,Bandyopadhyay,Tsai,Alexandre,MS}. In particular, Refs.~\cite{Fukushima,Ferrer, Bandyopadhyay} study the case of intense magnetic fields, where the lowest Landau Level (LLL) approximation can be used. Reference~\cite{Tsai} works the one-loop zero temperature case to all orders in the magnetic field and finds a general expression in terms of an integral over proper time parameters. No attempt to provide analytical results is made. In Ref.~\cite{Alexandre} the polarization tensor is computed both at finite temperature and field strength. The findings are applied to study magnetic field effects on the Debye screening. Reference~\cite{Ishikawa} expresses the one-loop polarization tensor as a sum over Landau levels and evaluates it using numerical methods. An analytical approach to the sum over Landau levels at zero temperature has been recently carried out in Ref.~\cite{new}. Magnetic corrections to the QCD equation of state in the hard thermal loop (HTL) and the LLL approximations, applied to the description of heavy-ion collisions have been considered in Refs.~\cite{Rath, Rath2}. Analytic results can be obtained in several limits of interest such as the HTL and LLL by considering different hierarchies for the fermion mass, thermal and magnetic scales~\cite{Bandyopadhyay,Karbstein}.

Nevertheless, in a thermo-magnetic medium, the gluon polarization tensor depends, in addition to the temperature $T$ and magnetic field strength $|eB|$, also on the square of its momentum components as well as on the fermion mass. The breaking of boost and rotational invariance introduced by thermal and magnetic effects, makes the polarization tensor to depend separately on the longitudinal and perpendicular components of the gluon momentum. The competition between these different energy scales produces a rich structure that can be better grasped if one resorts to analytic approximations that arise when considering given hierarchies of these energy scales. In this work we take on this task and consider the behavior of the gluon polarization tensor in vacuum and at high temperature for the case when the field is strong. The paper is organized as follows. In Sec.~\ref{sec:level2}, we compute the gluon polarization tensor in the presence of a strong, uniform and constant magnetic field  at zero and high temperature. For the latter case, working in the HTL approximation, we study two different hierarchies of scales:  $q^2\ll m_f^2 \ll T^2 $ and the the complementary case $m_f^2\ll q^2\ll T^2$, writing for each case the most general tensor structure and discussing its properties. We compute the magnetic modifications to the Debye mass, paying attention to the cases when longitudinal and the perpendicular components of the gluon momentum vanish at different rates.
 
In Sec.~\ref{sec:level3}, we summarize and discuss our results. We reserve for the appendices the calculation details for each of the regimes where the gluon polarization tensor is computed.

\section{\label{sec:level2}Thermo-magnetic gluon polarization tensor}

\begin{figure}[t!]
 \begin{center}
  \includegraphics[scale=0.6]{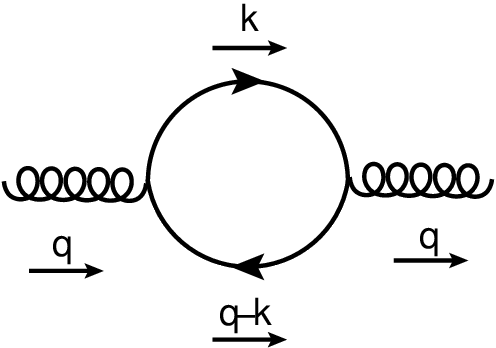}
 \end{center}
 \caption{One-loop diagram representing the gluon polarization tensor.}
 \label{fig1}
\end{figure}

We proceed to compute the gluon polarization tensor at one-loop order in the presence of a magnetic field, both in vacuum and in a thermal bath. In both cases we consider that the largest of all energy scales is the field strength. As we proceed to show, for the vacuum case, there is no need to establish a hierarchy of scales between the gluon momentum squared $q^2$ and the fermion mass squared $m_f^2$. However, for the thermal case, care has to be taken for the hierarchy between $q^2$ and $m_f^2$. Therefore, for the thermal case, we perform the computation in the two different regimes, namely $q^2 \ll m_f^2 \ll T^2$ and $m_f^2 \ll q^2 \ll T^2$. 

In general, the one-loop contribution to the gluon polarization tensor, depicted in Fig.~\ref{fig1}, is given by
\begin{equation}
\Pi_{ab}^{\mu \nu}=-\frac{1}{2}\int \frac{d^4k}{(2\pi)^4}\text{Tr}\{ igt_b\gamma^\nu iS_F(k)igt_a\gamma^\mu iS_F(k-q) \}.
\label{equ1}
\end{equation}
The factor $1/2$ accounts for the symmetry factor, which in the presence of the external magnetic field comes about given that the two contributing diagrams in Fig.~\ref{fig4}, with the opposite flow of charge, are not equivalent. Also, $g$ is the coupling constant, $S_F(k)$ is the quark propagator and $t_{a,b}$ are the generators of the color group. 
Since the quark anti-quark pair in the loop interact with the magnetic field, the quark propagator is modified from its vacuum expression and is written, omitting a trivial color factor, as
\begin{equation}
 S_F(x,x')=\Phi(x,x')\int \frac{d^4k}{(2\pi)^4} e^{-ik\cdot(x-x')}S(k),
 \label{equ2}
\end{equation}
where $\Phi(x,x')$ is called the \textit{Schwinger phase factor}. The latter accounts for the loss of Lorentz invariance in the presence of the magnetic field. For the present calculation, the phase factor can be {\it gauged away}~\cite{Chyi,vertex2} and we need to just work with the translationally invariant part of the fermion propagator. The latter is given by
\begin{align}
 iS(k)&=\int_0^\infty \frac{ds}{\cos(q_fBs)}e^{is(k_\parallel^2-k_\perp^2\frac{\tan(q_fBs)}{q_fBs}-m_f^2)} \nonumber \\
 &\times \Big \{ [\cos(q_fBs)+\gamma_1 \gamma_2 \sin(q_fBs)](m_f+\slashed{k}_\parallel) \nonumber \\ &-\frac{\slashed{k}_\perp}{\cos(q_fBs)}\big \},
 \label{equ3}
\end{align}
where $q_f$ is the the quark electric charge. Hereafter, we use the following notation for the parallel and perpendicular (with respect to the magnetic field) pieces of the scalar product of two four-vectors $a^\mu$ and $b^\mu$
\begin{align}
 (a\cdot b)_\parallel &= a_0b_0 - a_3b_3, \nonumber \\
 (a\cdot b)_\perp &= a_1b_1 +a_2b_2,
 \label{equ4}
\end{align}
such that 
\begin{equation}
 a\cdot b = (a\cdot b)_\parallel - (a\cdot b)_\perp.
\end{equation}

Gauge invariance requires that the gluon polarization tensor be transverse. However, the breaking of Lorentz symmetry makes this tensor to split into three transverse structures, such that the gluon polarization tensor can be written, omitting a trivial factor $\delta_{ab}$ coming from the color trace in Eq.~(\ref{equ1}), as~\cite{Hattori1} (see also Refs.~\cite{Ferrer,new, Aritra})
\begin{equation}
 \Pi^{\mu \nu}= P^\parallel \Pi_{\parallel}^{\mu \nu}+P^\perp \Pi_{\perp}^{\mu \nu}+P^0 \Pi_{0}^{\mu \nu},
 \label{equ5}
\end{equation}
where
\begin{align}
  \Pi_{\parallel}^{\mu\nu}  &= g^{\mu\nu}_\parallel-\frac{q^\mu_\parallel q^\nu_\parallel}{q^2_\parallel}, \nonumber \\
  \Pi_{\perp}^{\mu\nu}  &= g^{\mu\nu}_\perp-\frac{q^\mu_\perp q^\nu_\perp}{q^2_\perp}, \nonumber \\
  \Pi_{0}^{\mu \nu}&=g^{\mu\nu}-\frac{q^\mu q^\nu}{q^2}-(\Pi_{\parallel}^{\mu\nu}+\Pi_{\perp}^{\mu\nu}).
   \label{equ6}
\end{align}

Notice that the three tensor structures in Eq.~(\ref{equ6}) are orthogonal to each other, hence, their coefficients in Eq.~(\ref{equ5}) can be expressed as
\begin{align}
   P^{\parallel}  &=\Pi^{\parallel}_{\mu\nu} \ 
   \Pi^{\mu\nu}, \nonumber \\
  P^{\perp}  &=\Pi^{\perp}_{\mu\nu} \ \Pi^{\mu\nu}, \nonumber \\
   P^{0} &=\Pi^{0}_{\mu\nu} \ \Pi^{\mu\nu}.
   \label{equ7}
\end{align}

We now proceed to compute each of the coefficients in Eq.~(\ref{equ7}) in the strong field limit.

\subsection{Vacuum case, strong field approximation}

We now proceed to calculate the polarization tensor in the strong field limit, namely $|eB|\gg q^2,\ m_f^2$. The (photon) polarization tensor in this limit has been computed in Ref.~\cite{Fukushima} using Ritus' method. Here we work instead using Schwinger's proper time method. To implement this approximation, we work in the lowest Landau level (LLL) limit of the quark propagator, given by~\cite{MS}
\begin{equation}
 iS^{\text{LLL}}(k)=2ie^{-\frac{k_\perp^2}{|q_fB|}}\frac{\slashed{k}_\parallel+m_f}{k_\parallel^2-m_f^2}\mathcal{O}^\pm,
 \label{propLLL}
\end{equation}
where
\begin{equation}
 \mathcal{O}^\pm=\frac{1}{2}\left[1\pm i\gamma_1\gamma_2 \ \text{sign}(q_fB)\right].
 \label{o_pm}
\end{equation}

\begin{figure}[t!]
 \begin{center}
  \includegraphics[scale=0.5]{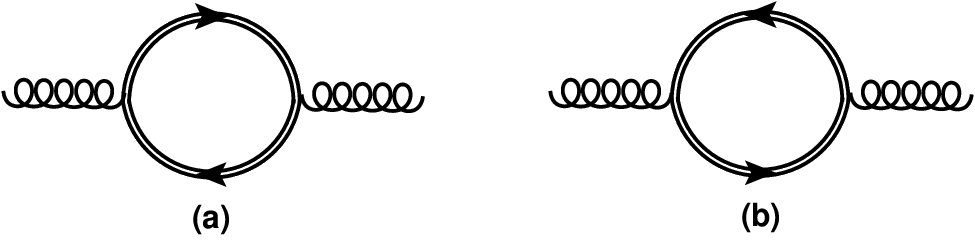}
 \end{center}
 \caption{One-loop diagrams for the gluon polarization tensor in the strong field limit, using the LLL.}
 \label{fig4}
\end{figure}

Figure~\ref{fig4} shows the diagrams contributing to the calculation. These represent one and the other possible electric charge flow direction within the loop, which, in the presence of the magnetic field have both to be accounted for. Using Eq.~(\ref{propLLL}), into the Eq.~(\ref{equ1}), the explicit expression for diagram $(a)$ in Fig.~(\ref{fig4}) is given by
\begin{align}
 i\Pi^{\mu\nu}_a&\!\!=\!\!\frac{-4g^2}{2}\sum_f \int \frac{d^2k_\perp}{(2\pi)^2}e^{-\frac{k_\perp^2}{|q_fB|}}e^{-\frac{(q_\perp-k_\perp)^2}{|q_fB|}} \nonumber \\
\!\!&\times\!\! \int \frac{d^2k_\parallel}{(2\pi)^2}\frac{\text{Tr}[\gamma^\nu(m_f-\slashed{k}_\parallel)\mathcal{O}^+\gamma^\mu(m_f-(\slashed{k}-\slashed{q})_\parallel)\mathcal{O}^+]}{[k_\parallel^2-m_f^2][(k-q)_\parallel^2-m_f^2]}.
 \label{poltenLLL}
\end{align}

The contribution from diagram $(b)$ in Fig.~\ref{fig4} is obtained by replacing $\mathcal{O}^+\to\mathcal{O}^-$. The polarization tensor is obtained by adding these two contributions, resulting in the  expression 
\begin{align}
 i\Pi^{\mu\nu}_a+i\Pi^{\mu\nu}_b&=-g^2\sum_f \int \frac{d^2k_\perp}{(2\pi)^2}e^{-\frac{k_\perp^2}{|q_fB|}}e^{-\frac{(q_\perp-k_\perp)^2}{|q_fB|}} \nonumber \\
 &\times \int \frac{d^2k_\parallel}{(2\pi)^2}\frac{1}{[k_\parallel^2-m_f^2][(k-q)_\parallel^2-m_f^2]} \nonumber \\
 \times& \{\text{Tr}[\gamma^\nu(m_f-\slashed{k}_\parallel)\mathcal{O}^+\gamma^\mu(m_f-(\slashed{k}-\slashed{q})_\parallel)\mathcal{O}^+]\nonumber \\
 +&\text{Tr}[\gamma^\nu(m_f-\slashed{k}_\parallel)\mathcal{O}^-\gamma^\mu(m_f-(\slashed{k}-\slashed{q})_\parallel)\mathcal{O}^-]\}.
 \label{twocontributions}
\end{align}
The explicit expressions for the traces are given by
\begin{align}
 &\text{Tr}[\gamma^\nu(m_f-\slashed{k}_\parallel)\mathcal{O}^\pm\gamma^\mu(m_f-(\slashed{k}-\slashed{q})_\parallel)\mathcal{O}^\pm]=\nonumber \\
 &m_f^2\text{Tr}[\gamma^\nu \mathcal{O}^\pm \gamma^\mu \mathcal{O}^\pm]
+\text{Tr}[\gamma^\nu \slashed{k}_\parallel \mathcal{O}^\pm \gamma^\mu (\slashed{k}-\slashed{q})_\parallel \mathcal{O}^\pm] \nonumber \\
&=m_f^2\text{Tr}[\gamma^\nu \mathcal{O}^\pm \gamma^\mu_\parallel]
+\text{Tr}[\gamma^\nu \slashed{k}_\parallel \mathcal{O}^\pm \gamma^\mu_\parallel (\slashed{k}-\slashed{q})_\parallel].
\label{traces}
\end{align}
Substituting Eqs.~(\ref{o_pm}) and~(\ref{traces}) into Eq.~(\ref{twocontributions}), we obtain
\begin{align}
 i\Pi^{\mu\nu}_a+i\Pi^{\mu\nu}_b&=-g^2\sum_f \int \frac{d^2k_\perp}{(2\pi)^2}e^{-\frac{k_\perp^2}{|q_fB|}}e^{-\frac{(q_\perp-k_\perp)^2}{|q_fB|}} \nonumber \\
 &\times \int \frac{d^2k_\parallel}{(2\pi)^2}\frac{1}{[k_\parallel^2-m_f^2][(k-q)_\parallel^2-m_f^2]} \nonumber \\
 &\times \big \{m_f^2\text{Tr}[\gamma^\nu (\mathcal{O}^+ +\mathcal{O}^-)\gamma^\mu_\parallel]\nonumber \\
 &+\text{Tr}[\gamma^\nu \slashed{k}_\parallel (\mathcal{O}^+ +\mathcal{O}^-) \gamma^\mu_\parallel (\slashed{k}-\slashed{q})_\parallel]\big \}\nonumber \\
&=-g^2\sum_f \int \frac{d^2k_\perp}{(2\pi)^2}e^{-\frac{k_\perp^2}{|q_fB|}}e^{-\frac{(q_\perp-k_\perp)^2}{|q_fB|}} \nonumber \\
 &\times \int \frac{d^2k_\parallel}{(2\pi)^2}\frac{1}{[k_\parallel^2-m_f^2][(k-q)_\parallel^2-m_f^2]}\nonumber \\
 &\times \big \{m_f^2\text{Tr}[\gamma^\nu \gamma^\mu_\parallel]+\text{Tr}[\gamma^\nu \slashed{k}_\parallel  \gamma^\mu_\parallel (\slashed{k}-\slashed{q})_\parallel]\big \}.
\end{align}
After integrating over the transverse components of the four-momentum, the expression for the polarization tensor becomes
\begin{align}
 i\Pi^{\mu\nu}_a+i\Pi^{\mu\nu}_b&=-g^2\sum_f\Big( \frac{\pi |q_fB|}{4\pi^2} \Big)e^{-\frac{q_\perp^2}{2|q_fB|}}\nonumber \\
   &\times\int \frac{d^2k_\parallel}{(2\pi)^2}
\frac{1}{[k_\parallel^2-m_f^2][(k-q)_\parallel^2-m_f^2]}\nonumber \\
 &\times \Big[(m_f^2-k_\parallel\cdot(k-q)_\parallel)g_\parallel^{\mu\nu}
 +k_\parallel^\mu(k-q)_\parallel^\nu\nonumber \\
 &+k_\parallel^\nu(k-q)_\parallel^\mu\Big].
 \label{PiParallel}
\end{align}

In order to compute the two dimensional integral over the parallel components, we use Feynman's parametrization. Thus, the denominator in Eq.~(\ref{PiParallel}) is written as
\begin{equation}
 \frac{1}{[k_\parallel^2-m_f^2][(k-q)_\parallel^2-m_f^2]}=\int_0^1  \frac{dx}{[l_\parallel^2-\Delta]^2},
 \label{FeynmanP}
\end{equation}
with
$
 l_\parallel=k_\parallel-(1-x)q_\parallel
$
and
$
 \Delta=m_f^2-x(1-x)q_\parallel^2.
$
Shifting the integration variable $k_\parallel
\to l_\parallel$ in Eq.~(\ref{PiParallel}), we obtain

\begin{align}
i\Pi^{\mu\nu}_a+i\Pi^{\mu\nu}_b&=-g^2\sum_f\Big( \frac{|q_fB|}{4\pi} \Big)e^{-\frac{q_\perp^2}{2|q_fB|}}\nonumber \\
&\int_0^1 dx \int \frac{d^2l_\parallel}{(2\pi)^2}\frac{ 1}{[l_\parallel^2-\Delta]^2}\nonumber \\
&\times \{ 2l_\parallel^\mu l_\parallel^\nu-g_\parallel^{\mu \nu}l_\parallel^2-2x(1-x)q_\parallel^\mu q_\parallel^\nu \nonumber \\
&+ g_\parallel^{\mu \nu}(m_f^2+x(1-x)q_\parallel^2) \},
 \label{PiFeynmanP}
\end{align}
where in the integrand we have already discarded linear terms in $l_\parallel$, since they give a vanishing contribution. In order to compute the momentum integrals, we recall the following well known relations
\begin{widetext}
\begin{align}
 \int \frac{d^dl_\parallel}{(2\pi)^d}\frac{1}{[l_\parallel^2-\Delta]^n}&=i\frac{(-1)^n}{(4\pi)^{d/2}}\frac{\Gamma(n-d/2)}{\Gamma(n)}\Bigg( \frac{1}{\Delta} \Bigg)^{n-d/2}, \nonumber \\
 \int \frac{d^dl_\parallel}{(2\pi)^d}\frac{l_\parallel^2}{[l_\parallel^2-\Delta]^n}&= i\frac{(-1)^{n-1}}{(4\pi)^{d/2}}\frac{d}{2} \frac{\Gamma(n-d/2-1)}{\Gamma(n)} \Bigg( \frac{1}{\Delta}\Bigg)^{n-d/2-1},\nonumber \\
 \int \frac{d^dl_\parallel}{(2\pi)^d}\frac{l_\mu l_\nu}{[l_\parallel^2-\Delta]^n}&=i\frac{(-1)^{n-1}}{(4\pi)^{d/2}}\frac{g_{\mu \nu}}{2} \frac{\Gamma(n-d/2-1)}{\Gamma(n)}
  \Bigg( \frac{1}{\Delta}\Bigg)^{n-d/2-1}.
 \label{PeskinExp}
\end{align}
\end{widetext}
Using these into Eq.~(\ref{PiFeynmanP}), we get
\begin{widetext}
\begin{align}
 i\Pi^{\mu\nu}_a+i\Pi^{\mu\nu}_b&=-g^2\sum_f \Big( \frac{|q_fB|}{4\pi} \Big)e^{-\frac{q_\perp^2}{2|q_fB|}} \frac{-i}{(4\pi)^{d/2}}\int_0^1 dx \Bigg\{ g_\parallel^{\mu \nu} \Gamma(1-d/2)\Bigg( \frac{1}{\Delta} \Bigg)^{1-d/2}- g_\parallel^{\mu \nu} \frac{d}{2}\Gamma(1-d/2)\Bigg( \frac{1}{\Delta} \Bigg)^{1-d/2}\nonumber \\
 &+2x(1-x)q_\parallel^\mu q_\parallel^\nu \Gamma(2-d/2)\Bigg( \frac{1}{\Delta} \Bigg)^{2-d/2}- g_\parallel^{\mu \nu}(m_f^2+x(1-x)q_\parallel^2)\Gamma(2-d/2)\Bigg( \frac{1}{\Delta} \Bigg)^{2-d/2}\Bigg\}.
 \label{Pi_d}
\end{align}
\end{widetext}
Equation~(\ref{Pi_d}) is apparently divergent when taking the limit $d\rightarrow 2$. To show that this is not the case, we first combine the terms proportional to the tensor structure $g_\parallel^{\mu \nu}$, namely, the first, second and fourth terms in Eq.~(\ref{Pi_d}), to obtain
\begin{widetext}
\begin{align}
&g_\parallel^{\mu \nu}\Bigg [\Gamma(1-d/2)\Bigg( \frac{1}{\Delta} \Bigg)^{1-d/2}-\frac{d}{2}\Gamma(1-d/2)\Bigg( \frac{1}{\Delta} \Bigg)^{1-d/2} -(m_f^2+x(1-x)q_\parallel^2)\Gamma(2-d/2)\Bigg( \frac{1}{\Delta} \Bigg)^{2-d/2}\Bigg]\nonumber \\
 &=g_\parallel^{\mu \nu} \Bigg [(1-d/2)\Gamma(1-d/2)\Delta -(m_f^2+x(1-x)q_\parallel^2)\Gamma(2-d/2)\Bigg]\Bigg( \frac{1}{\Delta} \Bigg)^{2-d/2}\nonumber \\
 &=g_\parallel^{\mu \nu} \Bigg [(m_f^2-x(1-x)q_\parallel^2) -(m_f^2+x(1-x)q_\parallel^2)\Bigg]\Gamma(2-d/2) \Bigg( \frac{1}{\Delta} \Bigg)^{2-d/2} \nonumber \\
 &=-2 g_\parallel^{\mu \nu} q_\parallel^2 
 \Gamma(2-d/2) \Bigg( \frac{1}{\Delta} \Bigg)^{2-d/2}
 \label{nodivergence}
\end{align}
\end{widetext}
Substituting Eq.~(\ref{nodivergence}) into Eq.~(\ref{Pi_d}), we get
\begin{widetext}
\begin{align}
 i\Pi^{\mu\nu}_a+i\Pi^{\mu\nu}_b&=-g^2\sum_f \Big( \frac{|q_fB|}{4\pi} \Big)e^{-\frac{q_\perp^2}{2|q_fB|}} \frac{i}{(4\pi)^{d/2}}\Gamma(2-d/2)\int_0^1 dx \Bigg( \frac{1}{\Delta} \Bigg)^{2-d/2}\nonumber \\
 &\times \left\{ g_\parallel^{\mu \nu}[-m_f^2+x(1-x)q_\parallel^2+m_f^2+x(1-x)q_\parallel^2]-2x(1-x)q_\parallel^\mu q_\parallel^\nu \right\}  \nonumber \\
 &=-g^2\sum_f \Big( \frac{|q_fB|}{4\pi} \Big)e^{-\frac{q_\perp^2}{2|q_fB|}} \frac{i}{(4\pi)^{d/2}}\Gamma(2-d/2) 2q_\parallel^2 \left\{g_\parallel^{\mu \nu}-\frac{q_\parallel^\mu q_\parallel^\nu}{q_\parallel^2}\right\} \int_0^1 dx \ x(1-x)\Bigg( \frac{1}{\Delta} \Bigg)^{2-d/2}.
 \label{finitePi}
\end{align}
\end{widetext}
Notice that Eq.~(\ref{finitePi}) is now free of divergences when taking the limit $d\to 2$. We thus obtain
\begin{eqnarray}
i\Pi^{\mu\nu}_a+i\Pi^{\mu\nu}_b&=&i g^2\sum_f \Big( \frac{|q_fB|}{8\pi^2} \Big)e^{-\frac{q_\perp^2}{2|q_fB|}} q_\parallel^2 \left\{g_\parallel^{\mu \nu}-\frac{q_\parallel^\mu q_\parallel^\nu}{q_\parallel^2}\right\}\nonumber \\
&\times& \int_0^1 dx \ \frac{-x(1-x)}{\Delta},
 \label{deq2Pi}
\end{eqnarray}
from where it is seen that the emerging tensor structure is equal to $\Pi_\parallel^{\mu \nu}$. Therefore, in the strong field limit, the coefficients $P^\perp$ and $P^0$ turn out to be equal to zero. 

\begin{figure}[t]
 \begin{center}
  \includegraphics[scale=0.65]{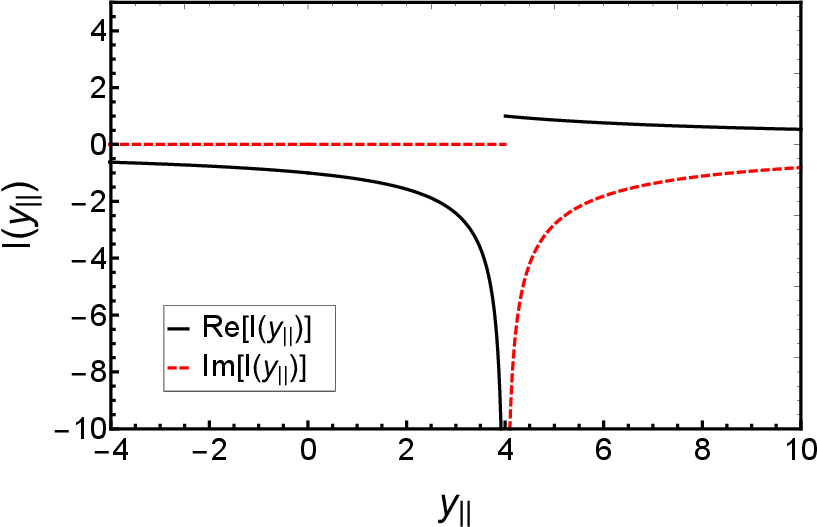}
 \end{center}
 \caption{Behavior of $I$ as a function of $y_\parallel$. Shown are the real (black) and imaginary (red) parts. This function presents a discontinuity at $y_\parallel=4$ that corresponds to the threshold for quark-antiquark production.}
 \label{fig5}
\end{figure}

The integral over the $x$ variable can be expressed in terms of the function 
\begin{align}
 I(y_\parallel)
 &= y_\parallel\int_0^1 dx \ \frac{x(1-x)}{-1+x(1-x)y_\parallel},
 \label{eqX}
\end{align}
where $y_\parallel\equiv q_\parallel^2/m_f^2$.
For the case $m_f=0$, $I=1$ and the gluon polarization tensor is given by
\begin{align}
 i\Pi^{\mu \nu}\left|_{m_f=0}\right.&=(i\Pi^{\mu\nu}_a+i\Pi^{\mu\nu}_b)\left|_{m_f=0}\right. \nonumber \\
 &=i g^2\left\{g_\parallel^{\mu \nu}-\frac{q_\parallel^\mu q_\parallel^\nu}{q_\parallel^2}\right\}\sum_f \Big( \frac{|q_fB|}{8\pi^2} \Big)e^{-\frac{q_\perp^2}{2|q_fB|}}.
\end{align}
This result coincides (albeit for the case of the photon polarization tensor) with the one found in Ref.~\cite{Fukushima} (see also Refs.~\cite{Ferrer,new}). Figure~\ref{fig5} shows the behavior of the function $I(y_\parallel)$ for the general case when $m_f^2\neq 0$. Notice that this function develops an imaginary part for $y_\parallel\geq 4$, corresponding to the threshold for quark-antiquark production. 

\subsection{Thermo-magnetic polarization tensor in the HTL and LLL approximations}

We now proceed to calculate the polarization tensor in the strong field limit within the HTL approximation. We consider the case where $|eB|\gg T^2$. At finite temperature, due to the presence of the vector $u^\mu$ that defines the medium's reference frame, the tensor $g^{\mu\nu}-q^\mu q^\nu/q^2$ splits into (three dimensional) transverse $\Pi_{T}^{\mu \nu}$ and longitudinal $\Pi_{L}^{\mu \nu}$ structures, such that
\begin{equation}
-g^{\mu \nu}+\frac{q^\mu q^\nu}{q^2} = \Pi_L^{\mu \nu}+\Pi_T^{\mu \nu},
\label{split}
\end{equation}
with
\begin{align}
    \Pi_{T}^{\mu\nu}  &= - g^{\mu\nu}+\frac{q^0}{\vec{q}^2}(q^\mu u^\nu+u^\mu q^\nu)-\frac{1}{\vec{q}^2}(q^\mu q^\nu+q^2u^\mu u^\nu),\nonumber\\
  \Pi_{L}^{\mu\nu}  &=-\frac{q^0}{\vec{q}^2}(q^\mu u^\nu+u^\mu q^\nu)
    +\frac{1}{\vec{q}^2}\Big[\frac{(q^0)^2}{q^2}q^\mu q^\nu+q^2u^\mu u^\nu\Big], 
\label{structuresLLLHTL}
\end{align}
where, in the medium's reference frame, $u^\mu=(1,0,0,0)$.

Notice that when working in the LLL, one can expect that the tensor structures in Eq.~(\ref{structuresLLLHTL}), together with $\Pi_\parallel^{\mu\nu}$, given by the first of Eqs.~(\ref{equ6}), span the most general expression for the polarization tensor for the case $|q_fB|\gg T^2$. Indeed, recall that when the gluon splits into a virtual quark-antiquark pair, and this occupies the same Landau level (the LLL in this case), the pair's transverse momentum vanishes, because the quark moves in the opposite direction than the antiquark around the field lines. From momentum conservation and for a finite gluon momentum, the motion of the virtual pair can therefore only happen along the direction of the external magnetic field. Since the quark/antiquark motion is thrust by the chromo-electric field, only a polarization  vector having a component along the external magnetic field can push the motion of the virtual pair~\cite{Hattori2}. The only polarization vector, and thus tensor, having a non-vanishing projection along the magnetic field is the parallel structure $\Pi_\parallel^{\mu\nu}$. In fact, from this argument, it can be expected that the transverse (to the magnetic field) polarization structure $\Pi_\perp^{\mu\nu}$ will be absent from the tensor basis. Also, when working in the limit $|eB|\gg T^2$, neither the virtual quark nor the antiquark have enough thermal energy to transit between Landau levels. Thus the motion of the virtual pair in the LLL would keep being along the external field. 

The above discussion applies to the situation where the thermo-magnetic system is not undergoing a collective motion characterized by a non-trivial medium’s flow velocity $u^\mu$. When this is not the case, an electric field  emerges, whose origin comes from boosting different system's elements to the medium rest frame. The simultaneous treatment of magnetic and electric field effects on medium's properties has been recently considered in the context of asymmetric collisions or event-by-event fluctuations in heavy-ion reactions~\cite{Ricardo}. Electromagnetic effects have also been studied using chiral kinetic theory with Landau level transitions induced by both boosts and rotations~\cite{Gao}. The subject is certainly of the utmost relevance for the physics of heavy-ion systems where one knows that the medium created in the reaction undergoes a collective expansion characterized by non-vanishing flow coefficients. Thus, we underline that our computation is valid only when the medium’s velocity is small such that the produced electric field cannot induce transitions between the LLL and higher Landau levels.

Concentrating only on the temperature dependent part of the polarization tensor $\widetilde{\Pi}^{\mu\nu}$, we can write
\begin{equation}
 \widetilde{\Pi}^{\mu \nu}= \widetilde{P}^L \Pi_L^{\mu \nu}+\widetilde{P}^T\Pi_T^{\mu \nu}.
 \label{tensorstructureBT}
\end{equation}
To find the coefficients  $\widetilde{P}^L$ and $\widetilde{P}^T$, we need to compute the corresponding projections onto $ \widetilde{\Pi}^{\mu \nu}$. This gives
\begin{eqnarray}
  \Pi_{\mu \nu}^L\widetilde{\Pi}^{\mu \nu}&=& \widetilde{P}^L, \nonumber \\
 \Pi_{\mu \nu}^T\widetilde{\Pi}^{\mu \nu}&=&  2\widetilde{P}^T.
 \label{coefficientsT}
\end{eqnarray}

In order to compute the contractions on the left-hand side of Eq.~(\ref{coefficientsT}), we follow a procedure similar to the one that lead to Eq.~(\ref{PiParallel}). This implies using that, in the strong field limit, transverse and parallel structures factorize, and also that temperature effects are obtained, in the Matsubara formalism, from the time-like component of the integration four-vector. Therefore all temperature effects are comprised to the parallel pieces of the integrals. Explicitly, within the Matsubara formalism, we transform the integrals to
Euclidean space by means of a Wick rotation, namely
\begin{equation}
\int \frac{d^4k}{(2\pi^4)} f(k) \rightarrow i T \sum_{n=-\infty}^{\infty} \int \frac{d^3k}{(2\pi^3)} f(i \tilde{\omega}_n,\vec{k}),
\end{equation}
where the integral over the time-like component of the
fer\-mi\-on momentum has been discretized and we introduced
the fer\-mi\-on Matsubara frequencies $\tilde{\omega}_n = (2n +1)\pi T$, $k_0=i \tilde{\omega}_n$ and $q_4=iq_0=\omega$. In order to obtain the coefficients, we need to compute the sum over Matsubara frequencies and the integral over $k_3$. 
Notice that after performing the sum, one gets two terms: one corresponding to the $T$-independent magnetized vacuum and another corresponding to the thermo-magnetic contribution. The magnetized vacuum term turns out, of course, to be equal to Eq.~(\ref{deq2Pi}) and therefore it comes entirely as the coefficient of $\Pi^{\mu\nu}_\parallel$. This is explicitly shown in Appendix~\ref{appB}. The temperature contribution is obtained from the terms proportional to the Fermi-Dirac distributions. For these terms, we work in two limits: (1) the external momentum $q$ is taken as the softest energy scale, thus we keep the quark mass finite and (2) the quark mass is the smallest energy scale and thus we keep the external momentum $q$ finite. 

\subsubsection{Case where $q^2\ll m_f^2\ll T^2$}\label{sec221}

When the external momentum $q$ is considered as the smallest energy scale, in the HTL approximation, we can neglect $q^2$ and $k\cdot q$ in each of the numerators. Thus, the explicit hierarchy of scales for this calculation is $q^2\ll m_f^2\ll T^2$. Performing the integral over $k_3$, and after analytical continuation back to Minkowski space, we get
\begin{align}
	\widetilde{P}^L&=\frac{g^2}{8\pi^2}\sum_f|q_fB| e^{-\frac{q_\perp^2}{2|q_fB|}}\Bigg [\frac{q_0^2(q_\perp^2+2q_3^2)}{\vec{q}^2 q^2}-1 \Bigg] \nonumber \\
	&\times \Bigg( \ln \Big( \frac{\pi^2 T^2}{m_f^2} \Big)-2\gamma_E \Bigg)
    \label{longfinal}
\end{align}
and
\begin{align}	
\widetilde{P}^T&=\frac{g^2}{8\pi^2}\sum_f|q_fB| e^{-\frac{q_\perp^2}{2|q_fB|}}\Bigg [\frac{q_{\perp}^2}{\vec{q}^2}\Bigg]\nonumber \\
&\times \Bigg ( \ln \Big( \frac{\pi^2 T^2}{m_f^2} \Big)-2\gamma_E \Bigg).
\label{transfinal}
\end{align}
The explicit calculations to obtain Eqs.~(\ref{longfinal}) and~(\ref{transfinal}) are shown in Appendix~\ref{appC}.
Notice that knowledge of the complete expression is important when discussing their   evolution properties under the renormalization group~\cite{Ayala:2018wux} and for the study of the thermo-magnetic effects on the Debye mass~\cite{Alexandre}.

\subsubsection{Case where $m_f^2\ll q^2\ll T^2$}\label{sec222}

Another way to compute the polarization tensor in the HTL approximation is to take into account the complementary hierarchy of scales $m_f^2\ll q^2\ll T^2$. After the integral over $k_3$ is performed and the analytical continuation to Minkowski space is done, we obtain
\begin{align}
    \widetilde{P}^L&=\frac{g^2}{64\pi^2}\sum_f |q_fB|e^{-\frac{q_\perp^2}{2|q_fB|}}\Bigg[ 2q_0^2+q^2+\frac{q_0^2q_3^2}{q^2} \Bigg] \nonumber \\
    &\times \Bigg( \frac{q_3^2}{\vec{q}^2q_\parallel^2}\Bigg)\Big(\frac{q_3}{T}\Big)
    \label{plhtl211}
\end{align}
and
\begin{equation}
   \widetilde{P}^T=-\frac{g^2}{64\pi^2}\sum_f |q_fB|e^{-\frac{q_\perp^2}{2|q_fB|}}\Bigg[ \frac{q_\perp^2 q_3^2}{q_\parallel^2 \vec{q}^2} \Bigg] \Big(\frac{q_3}{T}\Big).
    \label{pthtl2}
\end{equation}
The explicit calculations to obtain  Eqs.~(\ref{plhtl211}) and~(\ref{pthtl2}) are shown in Appendix~\ref{appD}. For this hierarchy of scales, we have already considered the massless limit $m_f\to 0$ and thus $q_3$ becomes the smallest energy scale. Also, notice that the coefficients $P^T, P^L$ vanish when we take the limit $q_3\rightarrow 0$. This shows that in this case, the matter contributions are all free of any infrared divergences.

\subsubsection{Magnetic corrections to the gluon Debye mass}\label{sec233}

If we now keep the fermion mass finite and work with the premises spelled out in subsection~\ref{sec221}, we can consider the matter contribution of Eqs.~(\ref{longfinal}) and~(\ref{transfinal}) to the dispersion relation. For this purpose, we need to take the limit when the three-mo\-men\-tum goes to zero. Nevertheless, notice that the result may be different depending on whether the parallel or perpendicular momentum component, with respect to the magnetic field, is taken first to zero. This behaviour is due to the breaking of the spatial isotropy and is the analog to the purely thermal case, where the limits when either $q_0$ or $|\vec{q}|$ goes first to zero, do not commute, due to the loss of Lorentz (boost) invariance. In the present case, the limits when either $q_3$ or $q_\perp$ go first to zero in general do not commute either. Let us explore the case when $q_\perp$ and $q_3$ goes to zero at different rates writing $q_3^2 = a \ q_\perp^2$. Thus, the matter contribution in Eqs.~(\ref{longfinal}) and~(\ref{transfinal}) can be written as
\begin{align}
	\widetilde{P}^L&=\frac{g^2}{8\pi^2}\sum_f|q_fB| e^{-\frac{q_\perp^2}{2|q_fB|}} \Bigg( \ln \Big( \frac{\pi^2 T^2}{m_f^2} \Big)-2\gamma_E \Bigg) \nonumber \\
    &\times  \Bigg [\frac{q_0^2(q_\perp^2+2aq_\perp^2)}{(1+a)q_\perp^2(q_0^2-(1+a)q_\perp^2)}-1 \Bigg]  \nonumber \\
    &=\frac{g^2}{8\pi^2}\sum_f|q_fB| e^{-\frac{q_\perp^2}{2|q_fB|}} \Bigg( \ln \Big( \frac{\pi^2 T^2}{m_f^2} \Big)-2\gamma_E \Bigg)  \nonumber \\
    &\times \Bigg [\frac{q_0^2(1+2a)}{(1+a)(q_0^2-(1+a)q_\perp^2)}-1 \Bigg] ,
    \label{longmatter}
\end{align}
\begin{align}	
\widetilde{P}^T&=\frac{g^2}{8\pi^2}\sum_f|q_fB| e^{-\frac{q_\perp^2}{2|q_fB|}} \Bigg( \ln \Big( \frac{\pi^2 T^2}{m_f^2} \Big)-2\gamma_E \Bigg) \nonumber \\
&\times \Bigg [\frac{q_{\perp}^2}{(1+a)q_\perp^2}\Bigg] \nonumber \\
&= \frac{g^2}{8\pi^2}\sum_f|q_fB| e^{-\frac{q_\perp^2}{2|q_fB|}} \Bigg( \ln \Big( \frac{\pi^2 T^2}{m_f^2} \Big)-2\gamma_E \Bigg)\nonumber \\
&\times \Bigg [\frac{1}{1+a}\Bigg].
\label{transmatter}
\end{align}

Notice that the behavior of $\widetilde{P}^T$ and $\widetilde{P}^L$ depend on $a$, giving rise to possibly different masses of the corresponding excitations when $q_\perp$ and $q_3$ go to zero at different rates. Thus, let us use Eqs.~(\ref{longmatter}) and~(\ref{transmatter}) to study the Debye mass, defined as the solution of
\begin{equation}
    [q^2-\widetilde{P}^{L,T} (q_0,q_\perp,q_3)]|_{\vec{q}=0}=0,
    \label{Debye}
\end{equation}
for $q_0=m_D$, in the different cases.
\begin{itemize}
\item $q_\perp\to 0$, $q_3$ finite:\\
In this case, Eqs.~(\ref{longmatter}) and~(\ref{transmatter}) become
\begin{eqnarray}
  \widetilde{P}^L&=&\frac{g^2}{8\pi^2}\sum_f|q_fB|\left(\frac{q_0^2}{q_\parallel^2}\right)\Bigg( \ln \Big( \frac{\pi^2 T^2}{m_f^2} \Big)-2\gamma_E \Bigg),\nonumber\\
  \widetilde{P}^T&=&0,
\label{case1}
\end{eqnarray}
which shows that transverse modes are not screened. If we now take $q_3\to 0$ in the first of Eqs.~(\ref{case1}), we see that the longitudinal mode develops a Debye mass given by
\begin{eqnarray}
(m_D^2)_L=\frac{g^2}{8\pi^2}\sum_f |q_fB|\Bigg(\ln\Big(\frac{\pi^2T^2}{m_f^2} \Big)-2\gamma_E \Bigg).
\label{masscase1}
\end{eqnarray}
\item $q_3\to 0$, $q_\perp$ finite:\\
In this other case, Eqs.~(\ref{longmatter}) and~(\ref{transmatter}) become
\begin{eqnarray}
  \widetilde{P}^L&=&\frac{g^2}{8\pi^2}\sum_f|q_fB|\Bigg( \ln \Big( \frac{\pi^2 T^2}{m_f^2} \Big)-2\gamma_E \Bigg)\nonumber\\
  &\times& \left(\frac{q_0^2}{q_0^2-q_\perp^2}-1\right), \nonumber \\
  \widetilde{P}^T&=&\frac{g^2}{8\pi^2}\sum_f|q_fB|\Bigg( \ln \Big( \frac{\pi^2 T^2}{m_f^2} \Big)-2\gamma_E \Bigg).
\label{case2}
\end{eqnarray}
If we now take $q_\perp \to 0$, $\widetilde{P}^L$ is equal to zero and this time it is the transverse modes which develop a Debye mass given by
\begin{eqnarray}
(m_D^2)_T=\frac{g^2}{8\pi^2}\sum_f|q_fB|\Bigg( \ln \Big( \frac{\pi^2 T^2}{m_f^2} \Big)-2\gamma_E \Bigg),
\label{masscase2}
\end{eqnarray}
whereas the longitudinal one is not screened.
Notice that the right-hand side of this expression coincides with the Debye mass for the longitudinal mode in the previous case given by Eq.~(\ref{masscase1}).
\item $q_3, \ q_\perp \to 0$ at the same rate:\\
In this last case, Eqs.~(\ref{longmatter}) and~(\ref{transmatter}) become
\begin{eqnarray}
  \widetilde{P}^L&=&\widetilde{P}^T=\frac{g^2}{16\pi^2}\sum_f|q_fB|\nonumber \\
  &\times& \Bigg( \ln \Big( \frac{\pi^2 T^2}{m_f^2} \Big)-2\gamma_E \Bigg),
\label{case3}
\end{eqnarray}
and both the longitudinal and transverse modes develop a Debye mass given by
\begin{eqnarray}
 (m_{D}^2)_L&=&(m_{D}^2)_T=\frac{g^2}{16\pi^2}\sum_f|q_fB|\nonumber \\
 &\times&\Bigg( \ln \Big( \frac{\pi^2 T^2}{m_f^2} \Big)-2\gamma_E \Bigg).
 \label{DebyemassLT}
\end{eqnarray}
Notice that when $q_3$ and $q_\perp$ vanish at the same rate, both modes are screened and the Debye mass of the longitudinal mode is equal to that of the transverse mode. 
\end{itemize}
For the three cases,
\begin{eqnarray}
(m_D^2)_L+(m_D^2)_T=\frac{g^2}{8\pi^2}\sum_f |q_fB|\Bigg(\ln\Big(\frac{\pi^2T^2}{m_f^2} \Big)-2\gamma_E \Bigg). \nonumber \\
\label{summasses}
\end{eqnarray}
We emphasize that the thermo-magnetic contribution to the polarization tensor can be expressed using only the orthogonal tensors $\Pi^{\mu\nu}_L$ and $\Pi^{\mu\nu}_T$, given by Eqs.~(\ref{structuresLLLHTL}). Indeed, one could have thought that the tensor $\Pi^{\mu\nu}_\parallel$ could also develop a thermo-magnetic coefficient. However, notice that since the projections $\Pi^{\mu\nu}_L\Pi_{\mu\nu}^\parallel$ and $\Pi^{\mu\nu}_T\Pi_{\mu\nu}^\parallel$ are non-vanishing, one can always express $\Pi^{\mu\nu}_\parallel$ as a linear combination of $\Pi^{\mu\nu}_L$ and $\Pi^{\mu\nu}_T$. Therefore it is sufficient to obtain the thermo-magnetic dependence of the coefficients of these latter, as has been expressed in Eq.~(\ref{coefficientsT}). In physical terms, this means that the matter corrections to the gluon dispersion properties can be calculated over a background of a magnetized vacuum.

\section{\label{sec:level3}Summary and discussion}

In this work we have computed the gluon polarization tensor in a thermo-magnetic medium. The computation has been performed including the magnetic field effects by means of Schwinger's proper time method. Although the vacuum polarization tensor for gauge fields has been previously studied in several other works (see for example Refs.~\cite{Dittrich,Hattori1, Hattori2, Fukushima, Tsai, Alexandre}), here we have analytically studied in detail the strong field limit at zero and high temperature. The latter case has been implemented within the HTL approximation.

For the $T=0$ case, we have computed the coefficients of the tensor structures describing the polarization tensor including the case where the quark mass is non-vanishing. We have shown that in the LLL, the only surviving structure corresponds to the tensor $\Pi_\parallel^{\mu\nu}$. The coefficient is given by an expression that develops an imaginary part, corresponding to the threshold for quark-antiquark pair production, appearing when the parallel gluon momentum squared is such that $q_\parallel^2= 4m_f^2$. For the case where the quark mass vanishes, this coefficient is real and equal to 1. To understand why it is that $q_\parallel^2$ dictates the threshold behaviour, we have noticed that when the gluon splits into a virtual quark-antiquark pair, in the same Landau level, the LLL in this case, the pair's total transverse momentum vanishes. This happens because the quark moves in the opposite direction than the antiquark around the field lines. From momentum conservation the motion of the virtual pair can  only happen along the direction of the external magnetic field and thus the kinematics is dictated by $q_\parallel^2$. This also helps understanding why the parallel tensor structure is the only one that survives.  Indeed, since the quark/antiquark motion is thrust by the chromo-electric field, only a polarization  vector having a component along the external magnetic field can push the motion of the virtual pair. The only polarization vector, and thus tensor, having a non-vanishing projection along the magnetic field is the parallel structure~\cite{Hattori2}.

For large $T$, we have also implemented the calculation in the LLL approximation. We have explicitly separated the magnetized vacuum and the thermo-magnetic contributions. This is particularly useful for a renormalization group analysis of the gluon polarization tensor, suited to extract the behavior of the strong coupling in a thermo-magnetic environment~\cite{Ayala:2018wux}.
We have worked within the hierarchy of scales $|eB|\gg T^2$ to make sure that thermal fluctuations do not induce transitions between higher Landau levels.
In this approximation, temperature and magnetic field effects factorize, due to the dimensional reduction in the LLL. This factorization is not possible when $T^2$ is larger than $|eB|$ and to describe such case, one would need to include in the calculation the contribution from other Landau levels. 

We analyzed the coefficients of the two tensor structures that appear at finite temperature and magnetic field and considered the cases $q^2\ll m_f^2\ll T^2$ and $m_f^2\ll q^2\ll T^2$. We showed that both cases are free from infrared divergences. In particular, we have obtained the thermo-magnetic behavior of the Debye mass from the gluon polarization tensor in the HTL and LLL approximations. However, since the magnetic field breaks rotational invariance, the result depends separately on the momentum components $q_3$ and $q_\perp$. Therefore one needs to consider distinct limits when either $q_3$ and $q_\perp$ go to zero at different rates. We distinguish three different cases: When $q_\perp$ ($q_3$) goes first to zero, only the longitudinal (transverse) mode develops a Debye mass. However, when both $q_\perp$ and $q_3$ go to zero at the same rate, both modes develop the same Debye mass. Consequences of these different screening patterns will be explored in the future and reported elsewhere.

\begin{acknowledgments}
This work was supported by Consejo Nacional de Ciencia y Tecnolog\'ia grant number 256494, by UNAM-DGAPA-PAPIIT grant number IG100219, by University of Cape Town, by Fondecyt (Chile) grant numbers 1170107, 1190192, Conicyt/PIA/Basal (Chile) grant number FB0821. R. Zamora acknowledges support from FONDECYT (Chile) under grant No. 1200483. D. M. acknowledges support from a PAPIIT-DGAPA-UNAM fellowship.
\end{acknowledgments}

\appendix
\begin{widetext}
\section{Vacuum contribution in the HTL and LLL approximations}\label{appB}

 Here we show that the vacuum contribution, after the sum over Matsubara frequencies, is equal to the integral over $k_0$ in Eq.~(\ref{PiParallel}). This means that the vacuum contribution at $T,\ eB \neq 0$ in the LLL and HTL approximations yields the same result as the case for $T=0$ in the presence of a magnetic field in the LLL approximation. We begin with Eq.~(\ref{PiParallel})
\begin{align}
 i\Pi^{\mu\nu}_a+i\Pi^{\mu\nu}_b&=g^2\sum_f\Big( \frac{\pi |q_fB|}{4\pi^2} \Big)e^{-\frac{q_\perp^2}{2|q_fB|}}\int \frac{d^2k_\parallel}{(2\pi)^2}
\frac{1}{[k_\parallel^2-m_f^2][(k-q)_\parallel^2-m_f^2]}\nonumber \\
 &\times \Big[(m_f^2-k_\parallel\cdot(k-q)_\parallel)g_\parallel^{\mu\nu}
 +k_\parallel^\mu(k-q)_\parallel^\nu+k_\parallel^\nu(k-q)_\parallel^\mu\Big] \nonumber \\
 &=g^2\sum_f\Big( \frac{\pi |q_fB|}{4\pi^2} \Big)e^{-\frac{q_\perp^2}{2|q_fB|}}\int \frac{d^2k_\parallel}{(2\pi)^2}
\frac{1}{[k_\parallel^2-m_f^2][(k-q)_\parallel^2-m_f^2]}\nonumber \\
 &\times \Big[(m_f^2+k_3^2-k_3q_3+k_0q_0-k_0^2)g_\parallel^{\mu\nu}+2\big(k_3^2-k_3(q_0+q_3)\nonumber\\
 &+k_0(2k_3-q_3-q_0)+k_0^2 \big)\Big] \nonumber \\
 &=g^2\sum_f\Big( \frac{\pi |q_fB|}{4\pi^2} \Big)e^{-\frac{q_\perp^2}{2|q_fB|}}\int \frac{d^2k_\parallel}{(2\pi)^2}
\frac{1}{[k_\parallel^2-m_f^2][(k-q)_\parallel^2-m_f^2]}\nonumber \\
 &\times \Big[(m_f^2+k_3^2-k_3q_3)g_\parallel^{\mu\nu}+2(k_3^2-k_3(q_3+q_0))
 \nonumber \\
 &+k_0(q_0 g_\parallel^{\mu\nu}+2(k_3-q_3-q_0)) +k_0^2(-g_\parallel^{\mu\nu}+2) \Big].
  \end{align}
Changing to Euclidean space by means of a Wick rotation and working in the imaginary time formalism with $q_0=i \omega$ and $k_0=i\widetilde{\omega}_n$, the polarization tensor becomes
\begin{align}
 i\Pi^{\mu\nu}_a+i\Pi^{\mu\nu}_b&=i g^2\sum_f\Big(\frac{\pi |q_fB|}{4\pi^2}  \Big)e^{-\frac{q_\perp^2}{2|q_fB|}} T \sum_{n=-\infty}^{\infty}\int \frac{dk_3}{(2\pi)^2}\Bigg[
\frac{(m_f^2+k_3^2-k_3q_3)g_\parallel^{\mu\nu}+2(k_3^2-k_3(q_3+i \omega))}{[\tilde{\omega}_n^2+k_3^2+m_f^2][(\tilde{\omega}_n-\omega)^2+(k_3-q_3)^2+m_f^2]} \nonumber \\
&-\frac{(i\widetilde{\omega}_n)(i\omega g_\parallel^{\mu\nu}+2(k_3-q_3-i\omega))}{[\tilde{\omega}_n^2+k_3^2+m_f^2][(\tilde{\omega}_n-\omega)^2+(k_3-q_3)^2+m_f^2]} +\frac{(i\widetilde{\omega}_n)^2(-g_\parallel^{\mu\nu}+2)}{[\tilde{\omega}_n^2+k_3^2+m_f^2][(\tilde{\omega}_n-\omega)^2+(k_3-q_3)^2+m_f^2]}\Bigg].
\label{tensorsum}
\end{align}
In order to compute the sum over Matsubara frequencies, we use the following expressions
\begin{align}
I_0 &= T \sum_{n=-\infty}^{\infty}\frac{1}{[\tilde{\omega}_n^2+k_3^2+m_f^2][(\tilde{\omega}_n-\omega)^2+(k_3-q_3)^2+m_f^2]}\nonumber\\ &=- \sum_{s_1,s_2 = \pm1} \frac{s_1s_2}{4E_1 E_2} \Bigr[\frac{1-\tilde{f}(s_1 E_1)-\tilde{f}(s_2 E_2)}{i \omega -s_1 E_1-s_2 E_2} \Bigl], \nonumber \\
I_1 &= T \sum_{n=-\infty}^{\infty}\frac{i \widetilde{\omega}_n}{[\tilde{\omega}_n^2+k_3^2+m_f^2][(\tilde{\omega}_n-\omega)^2+(k_3-q_3)^2+m_f^2]}\nonumber\\ &= -\sum_{s_1,s_2 = \pm1} \frac{s_1s_2 (s_1 E_1)}{4E_1 E_2} \Bigr[\frac{1-\tilde{f}(s_1 E_1)-\tilde{f}(s_2 E_2)}{i \omega -s_1 E_1-s_2 E_2} \Bigl], \nonumber \\
I_3 &= T \sum_{n=-\infty}^{\infty}\frac{(i \widetilde{\omega}_n)^2}{[\tilde{\omega}_n^2+k_3^2+m_f^2][(\tilde{\omega}_n-\omega)^2+(k_3-q_3)^2+m_f^2]}\nonumber\\ &=-\sum_{s_1,s_2 = \pm1} \frac{s_1s_2 (s1 E_1)(s_2E_2-i \omega)}{4E_1 E_2} \Bigr[\frac{1-\tilde{f}(s_1 E_1)-\tilde{f}(s_2 E_2)}{i \omega -s_1 E_1-s_2 E_2} \Bigl].
\label{mastersums}
\end{align}
We now substitute the vacuum pieces in Eqs.~(\ref{mastersums}) into Eq.~(\ref{tensorsum}). This means that we only consider the terms independent of the distribution functions $\tilde{f}$. We obtain
\begin{align}
 i\Pi^{\mu\nu}_a+i\Pi^{\mu\nu}_b&=-i g^2\sum_f\Big(\frac{\pi |q_fB|}{4\pi^2}  \Big)e^{-\frac{q_\perp^2}{2|q_fB|}} \nonumber \\
   &\times\int \frac{dk_3}{(2\pi)^2}\Bigg[
\big[(m_f^2+k_3^2-k_3q_3)g_\parallel^{\mu\nu}+2(k_3^2-k_3(q_3+i \omega))\big] \frac{1}{4E_1E_2} \Big(\frac{1}{i \omega - E_1- E_2}- \frac{1}{i\omega+ E_1+E_2}\Big)
\nonumber \\
&-(i\omega g_\parallel^{\mu\nu}+2(k_3-q_3-i\omega))  \frac{E_1}{4E_1E_2}\Big(\frac{1}{i \omega - E_1- E_2}- \frac{1}{i\omega+ E_1+E_2}\Big) \nonumber \\
&+\frac{(-g_\parallel^{\mu\nu}+2)}{4E_1E_2}\Bigg(-E_1E_2\Big(\frac{1}{i \omega-E_1-E_2}- \frac{1}{i \omega+E_1+E_2}\Big)E_1(i \omega) \Big(\frac{1}{i \omega-E_1-E_2}+ \frac{1}{i \omega+E_1+E_2}\Big)\Bigg)
\Bigg].\nonumber\\
\label{finalsumando}
\end{align}
On the other hand, computing the integral over $k_0$ in Eq.~(\ref{PiParallel}), one gets
\begin{align}
i\Pi^{\mu\nu}_a+i\Pi^{\mu\nu}_b&= g^2\sum_f\Big(\frac{\pi |q_fB|}{4\pi^2}  \Big)e^{-\frac{q_\perp^2}{2|q_fB|}}\int \frac{d^2k_\parallel}{(2\pi)^2}
\frac{1}{[k_\parallel^2-m_f^2+i \epsilon][(k-q)_\parallel^2-m_f^2+i \epsilon]} \nonumber \\
&\times \Big[(m_f^2+k_3^2-k_3q_3)g_\parallel^{\mu\nu}+2(k_3^2-k_3(q_3+q_0))+k_0(q_0 g_\parallel^{\mu\nu}+2(k_3-q_3-q_0)) +k_0^2(-g_\parallel^{\mu\nu}+2) \Big] \nonumber \\
 &= g^2\sum_f\Big(\frac{\pi |q_fB|}{4\pi^2}  \Big)e^{-\frac{q_\perp^2}{2|q_fB|}} \int \frac{dk_3}{(2\pi)}\int \frac{dk_0}{(2\pi)}\nonumber \\
 &\times \frac{1}{[k_0-E_1+i \epsilon][k_0+E_1-i \epsilon]} \frac{1}{[k_0-(q_0+E_2)+i \epsilon][k_0-(q_0-E_2)-i \epsilon]}\nonumber \\
   &\times
\Big[(m_f^2+k_3^2-k_3q_3)g_\parallel^{\mu\nu}+2(k_3^2-k_3(q_3+q_0)) k_0(q_0 g_\parallel^{\mu\nu}+2(k_3-q_3-q_0)) +k_0^2(-g_\parallel^{\mu\nu}+2) \Big],
\end{align}
Integrating over $k_0$ in the complex plane, where we take a closed path that encloses either the upper or the lower half-plane, since, in either case, one always encloses two of the poles, the result is
\begin{align}
 i\Pi^{\mu\nu}_a+i\Pi^{\mu\nu}_b&=-i g^2\sum_f\Big(\frac{\pi |q_fB|}{4\pi^2}  \Big)e^{-\frac{q_\perp^2}{2|q_fB|}} \nonumber \\
   &\times\int \frac{dk_3}{(2\pi)^2}\Bigg[
\big[(m_f^2+k_3^2-k_3q_3)g_\parallel^{\mu\nu}+2(k_3^2-k_3(q_3+q_0))\big] \frac{1}{4E_1E_2} \Big(\frac{1}{q_0 - E_1- E_2}- \frac{1}{q_0+ E_1+E_2}\Big)
\nonumber \\
&-\big(q_0 g_\parallel^{\mu\nu}+2(k_3-q_3-q_0)\big) \frac{E_1}{4E_1E_2}\Big(\frac{1}{q_0 - E_1- E_2}- \frac{1}{q_0+ E_1+E_2}\Big) \nonumber \\
&+\frac{(-g_\parallel^{\mu\nu}+2)}{4E_1E_2}\Bigg(-E_1E_2\Big(\frac{1}{q_0-E_1-E_2}- \frac{1}{q_0+E_1+E_2}\Big)+E_1q_0 \Big(\frac{1}{q_0-E_1-E_2}+ \frac{1}{q_0+E_1+E_2}\Big)\Bigg)
\Bigg].
\label{finalintegrando}
\end{align}
Comparing Eq.~(\ref{finalintegrando}) and Eq.~(\ref{finalsumando}), we see that the result is the same upon the analytical continuation $i\omega\to q_0$.

\section{Integrals for the polarization tensor in the LLL and HTL approximations ($q^2\ll m_f^2\ll T^2$)}\label{appC}

Here we show the explicit steps that lead to the results for the matter contributions in Eqs.~(\ref{longfinal}) and~(\ref{transfinal}). First, we introduce temperature effects into the corresponding projections onto $\widetilde{\Pi}^{\mu \nu}$, Eq. (\ref{coefficientsT}), using the Matsubara formalism of thermal field theory, obtaining
    
\begin{align}
 \widetilde{P}^L&=-g^2T\sum_{n=-\infty}^{\infty}\sum_f\Big( \frac{\pi |q_fB|}{4\pi^2} \Big)e^{-\frac{q_\perp^2}{2|q_fB|}} \int \frac{dk_3}{(2\pi)}\nonumber \\
 &\times \left\{  
\frac{1}{[\tilde{\omega}_n^2+k_3^2+m_f^2][(\tilde{\omega}_n-\omega)^2+(k_3-q_3)^2+m_f^2]} \right\} \nonumber  \\
 &\times \left\{  \Big[4  \tilde{\omega}_n( -\tilde{\omega}_n \omega -k_3 q_3 ) +2\tilde{\omega}_n (\omega_n^2+q_3^2) + 2 \omega (\tilde{\omega}_n^2+k_3^2+m_f^2) \Big] \frac{\omega}{\vec{q}^2}
 \nonumber \right.\\ 
 &+ \left. \Big[2(\tilde{\omega}_n^2 \omega^2+k_3^2 q_3^2+2\tilde{\omega}_n \omega k_3 q_3) - (\omega_n^2+q_3^2) (\tilde{\omega}_n \omega +k_3 q_3+m_f^2) \Big] \frac{\omega^2}{\vec{q}^2 q^2}
 \nonumber \right. \\
 &+ \left. \Big[\tilde{\omega}_n^2+2\tilde{\omega}_n \omega -k_3^2 +\tilde{\omega}_n \omega +k_3q_3-m_f^2 \Big] \frac{q^2}{\vec{q}^2}\right\}\label{paralelterm2}
\end{align}  
and   
\begin{equation}    
 \widetilde{P}^T=-g^2T\sum_{n=-\infty}^{\infty}\sum_f\Big( \frac{\pi |q_fB|}{4\pi^2} \Big)e^{-\frac{q_\perp^2}{2|q_fB|}}\frac{q_\perp^2}{\vec{q}^2} \int \frac{dk_3}{(2\pi)} \left\{ 
\frac{(-\tilde{\omega}_n^2+k_3^2-m_f^2-\tilde{\omega_n}\omega -k_3q_3)}{[\tilde{\omega}_n^2+k_3^2+m_f^2][(\tilde{\omega}_n-\omega)^2+(k_3-q_3)^2+m_f^2]} \right\}. \label{paralelterm1}
\end{equation}  
We notice that since in the HTL approximation terms proportional to $\tilde{\omega}_n$ and to $k_3$ in the numerators do not contribute, the calculation of Eqs.~(\ref{paralelterm2}) and~(\ref{paralelterm1}) involves only two kinds of sums over the Matsubara frequencies. These are explicitly given by~\cite{LeBellac}
\begin{equation}
 T \sum_{n=-\infty}^{\infty}\frac{1}{[(\tilde{\omega}_n-\omega)^2+(k_3-q_3)^2+m_f^2]} = \frac{1}{2E_2}(1-2\tilde{f}(E_2))\label{suma1},
\end{equation}
and
\begin{equation}
 T \sum_{n=-\infty}^{\infty}\frac{1}{[\tilde{\omega}_n^2+k_3^2+m_f^2][(\tilde{\omega}_n-\omega)^2+(k_3-q_3)^2+m_f^2]}= \sum_{s_1,s_2 = \pm1} \frac{-s_1s_2}{4E_1 E_2} \Bigr[\frac{1-\tilde{f}(s_1 E_1)-\tilde{f}(s_2 E_2)}{i \omega -s_1 E_1-s_2 E_2} \Bigl] \label{suma2},
\end{equation}
with $E_1^2=k_3^2+m_f^2$, $E_2^2=(k_3-q_3)^2+m_f^2$ and $\tilde{f}(x)=1/(e^{x/T}+1)$. Using Eqs.~(\ref{suma1}) and~(\ref{suma2}) and continuing within the HTL approximation, we notice that the external momentum $q$ is a soft energy scale compared with the temperature and quark mass, and therefore we can neglect $q^2$ and $k\cdot q$ in each of the numerators. We can therefore compute all the sums appearing in Eqs.~(\ref{paralelterm2}) and~(\ref{paralelterm1}), ignoring the vacuum contributions, we obtain
\begin{equation}
\widetilde{P}^L=-g^2\sum_f\Big( \frac{|q_fB|}{8\pi^2} \Big)e^{-\frac{q_\perp^2}{2|q_fB|}} \left\{\int dk_3 \left( \frac{2\tilde{f}(E_1)}{E_1} \right) \left( \frac{\omega^4+\vec{q}^4}{\vec{q}^2 q^2} \right) 
+ \int dk_3 \left( -\frac{2\tilde{f}(E_1)}{E_1} \right) \left( \frac{q_3^2\omega^2 -\vec{q}^4}{\vec{q}^2 q^2} \right) \right\}
\end{equation}
and
\begin{equation}
\widetilde{P}^T=-\frac{g^2}{2}\sum_f\Big( \frac{|q_fB|}{4\pi^2} \Big)e^{-\frac{q_\perp^2}{2|q_fB|}}\frac{q_\perp^2}{\vec{q}^2} \int dk_3 \left(-\frac{2 \tilde{f}(E_1)}{E_1} \right).
\end{equation}

We now turn to compute the matter contribution. For this we require an expression for
\begin{equation}
\int_{-\infty}^\infty dk_3 \frac{\tilde{f(E_1)}}{E_1} = \int_{-\infty}^\infty dk_3\frac{1}{(k_3^2+m_f^2)^{1/2}} \frac{1}{e^{\sqrt{(k_3^2+m_f^2)}/T}+1}.
\label{mattercontribution}
\end{equation}
Using the general expression from  Ref.~\cite{Kapusta} 
\begin{equation}
f_n(\tilde{y})=\frac{1}{\Gamma(n)} \int_0^{\infty}\frac{dx x^{n-1}}{\sqrt{x^2+\tilde{y}^2}} \frac{1}{e^{\sqrt{x^2+\tilde{y}^2}}+1},
\label{Kapustafn}
\end{equation}
we identify Eq.~(\ref{mattercontribution}) as corresponding to the case with $n=1$ and $\tilde{y}=m_f/T$. In the limit where  $\tilde{y}$ is small, Eq.~(\ref{Kapustafn}) becomes
 \begin{equation}
  f_1(m_f/T)=-\frac{1}{2} \ln\left(\frac{m_f}{\pi T}\right) - \frac{1}{2} \gamma_E + \ldots,
   \end{equation}
therefore we obtain
\begin{eqnarray}
 \!\!\!\!\!\! \int_{-\infty}^\infty dk_3 \frac{\tilde{f(E_1)}}{E_1} = -\left(\ln\left(\frac{m_f}{\pi T}\right) +\gamma_E  \right). \nonumber \\
 \end{eqnarray}
Using the above expression, we get
\begin{equation}
	\widetilde{P}^L=-\frac{g^2}{8\pi^2}\sum_f|q_fB| e^{-\frac{q_\perp^2}{2|q_fB|}} \Bigg [\frac{-\omega^2(q_\perp^2+2q_3^2)}{\vec{q}^2 q^2}-1 \Bigg]\Bigg ( \ln \Big( \frac{m_f^2}{\pi^2 T^2} \Big)+2\gamma_E \Bigg)
    \label{longfinalC}
\end{equation}
and
\begin{align}	
\widetilde{P}^T&=-\frac{g^2}{8\pi^2}\sum_f|q_fB| e^{-\frac{q_\perp^2}{2|q_fB|}}\Bigg [\frac{q_{\perp}^2}{\vec{q}^2}\Bigg]\Bigg( \ln \Big( \frac{m_f^2}{\pi^2 T^2} \Big)+2\gamma_E \Bigg).
\label{transfinalC}
\end{align}
Finally, to obtain Eqs.~(\ref{longfinal}) and~(\ref{transfinal}) we perform the analytical continuation $i\omega\to q_0$, back to Minkowski space.

\section{Integrals for the polarization tensor in the LLL and HTL approximations ($m_f^2\ll q^2\ll T^2$)}\label{appD}

We now compute the matter contribution in the same fashion as~\ref{appC} but with the new hierarchy of scales $m_f^2\ll q^2\ll T^2$. We show the explicit steps that lead to the results for the matter contributions in Eqs.~(\ref{plhtl211}) and~(\ref{pthtl2}), where the internal momentum $k$ is the largest energy scale and the quark mass $m_f$ is the smallest one. First, we introduce temperature effects into the corresponding projections onto $\widetilde{\Pi}^{\mu \nu}$, Eq. (\ref{coefficientsT}), using the Matsubara formalism of thermal field theory, obtaining
\begin{align}
 \widetilde{P}^L&=-g^2T\sum_{n=-\infty}^{\infty}\sum_f\Big( \frac{\pi |q_fB|}{4\pi^2} \Big)e^{-\frac{q_\perp^2}{2|q_fB|}} \int \frac{dk_3}{(2\pi)}\nonumber \\
 &\times \left\{  
\frac{1}{[\tilde{\omega}_n^2+k_3^2+m_f^2][(\tilde{\omega}_n-\omega)^2+(k_3-q_3)^2+m_f^2]} \right\} \nonumber  \\
 &\times \left\{  \Big[4  \tilde{\omega}_n( -\tilde{\omega}_n \omega -k_3 q_3 ) +2\tilde{\omega}_n (\omega_n^2+q_3^2) + 2 \omega (\tilde{\omega}_n^2+k_3^2+m_f^2) \Big] \frac{\omega}{\vec{q}^2}
 \nonumber \right.\\ 
 &+ \left. \Big[2(\tilde{\omega}_n^2 \omega^2+k_3^2 q_3^2+2\tilde{\omega}_n \omega k_3 q_3) \nonumber \right. \\ &- \left. (\omega_n^2+q_3^2) (\tilde{\omega}_n \omega +k_3 q_3+m_f^2) \Big] \frac{\omega^2}{\vec{q}^2 q^2}
 \nonumber \right. \\
 &+ \left. \Big[\tilde{\omega}_n^2+2\tilde{\omega}_n \omega -k_3^2 +\tilde{\omega}_n \omega +k_3q_3-m_f^2 \Big] \frac{q^2}{\vec{q}^2}\right\}\label{paraleltermd}
\end{align}   
and   
\begin{equation}    
 \widetilde{P}^T=-g^2T\sum_{n=-\infty}^{\infty}\sum_f\Big( \frac{\pi |q_fB|}{4\pi^2} \Big)e^{-\frac{q_\perp^2}{2|q_fB|}}\frac{q_\perp^2}{\vec{q}^2} \int \frac{dk_3}{(2\pi)} \left\{ 
\frac{(-\tilde{\omega}_n^2+k_3^2-m_f^2-\tilde{\omega_n}\omega -k_3q_3)}{[\tilde{\omega}_n^2+k_3^2+m_f^2][(\tilde{\omega}_n-\omega)^2+(k_3-q_3)^2+m_f^2]} \right\}. \label{paralelterm1d}
\end{equation} 
We notice that since in the HTL approximation terms proportional to $m_f^2$, $\tilde{\omega}_n$, and $k_3$ in the numerators do not contribute, the calculation of Eqs.~(\ref{paraleltermd}) and~(\ref{paralelterm1d}) involves only two kinds of sums over the Matsubara frequencies, 
\begin{equation}
 T \sum_{n=-\infty}^{\infty}\frac{1}{[\tilde{\omega}_n^2+k_3^2+m_f^2][(\tilde{\omega}_n-\omega)^2+(k_3-q_3)^2+m_f^2]}= \sum_{s_1,s_2 = \pm1} \frac{-s_1s_2}{4E_1 E_2} \Bigr[\frac{1-\tilde{f}(s_1 E_1)-\tilde{f}(s_2 E_2)}{i \omega -s_1 E_1-s_2 E_2} \Bigl] \equiv \chi_0 \label{suma2ddd},
\end{equation}
and
\begin{equation}
 T \sum_{n=-\infty}^{\infty}\frac{\tilde{\omega}^2_n}{[\tilde{\omega}_n^2+k_3^2+m_f^2][(\tilde{\omega}_n-\omega)^2+(k_3-q_3)^2+m_f^2]}= \sum_{s_1,s_2 = \pm1} \frac{s_1s_2 E_1}{4E_2} \Bigr[\frac{1-\tilde{f}(s_1 E_1)-\tilde{f}(s_2 E_2)}{i \omega -s_1 E_1-s_2 E_2} \Bigl] \equiv \chi_1 \label{sumad},
\end{equation}
with $E_1^2=k_3^2+m_f^2$, $E_2^2=(k_3-q_3)^2+m_f^2$ and
\begin{eqnarray}
\tilde{f}(x)=\frac{1}{(e^{x/T}+1)}.
\end{eqnarray}
We observe that Eqs.~(\ref{suma2ddd}) and~(\ref{sumad}) are related by $\chi_1=-E_1^2\chi_0$. Therefore, we should only find an expression for the sum in Eq.~(\ref{sumad}).
In this approximation ($k>q>m_f$) Eq.~(\ref{sumad}) is given by
\begin{equation}
 T \sum_{n=-\infty}^{\infty}\frac{\tilde{\omega}^2_n}{[\tilde{\omega}_n^2+k_3^2+m_f^2][(\tilde{\omega}_n-\omega)^2+(k_3-q_3)^2+m_f^2]}= \frac{-q_3}{2 q^2_{\parallel}} \Bigl[\frac{q_3 e^{k_3/T}}{T(e^{k_3/T}+1)^2}+\frac{1}{2} \frac{q_3^2 e^{k_3/T}(e^{k_3/T}-1)}{T^2 (e^{k_3/T}+1)^3}+...\Bigr] \label{sumad2},
\end{equation}
and thus, Eqs.~(\ref{paraleltermd}) and~(\ref{paralelterm1d}) take the following form
\begin{align}
 \widetilde{P}^L&=-g^2T\sum_{n=-\infty}^{\infty}\sum_f\Big( \frac{\pi |q_fB|}{4\pi^2} \Big)e^{-\frac{q_\perp^2}{2|q_fB|}} \int \frac{dk_3}{(2\pi)}  
\frac{1}{[\tilde{\omega}_n^2+k_3^2+m_f^2][(\tilde{\omega}_n-\omega)^2+(k_3-q_3)^2+m_f^2]}  \nonumber  \\
 &\times \left\{ \Bigl[-4\tilde{\omega}_n^2 \omega + 2 \omega (\tilde{\omega}_n^2+k_3^2)\Bigr]\frac{\omega}{\vec{q}^2}
 + \Bigl[2(\tilde{\omega}_n^2 \omega^2+k_3^2q_3^2)\Bigr] \frac{\omega^2}{\vec{q}^2 q^2}
 + \Bigl[\tilde{\omega}_n^2-k_3^2\Bigr] \frac{q^2}{\vec{q}^2}\right\}\label{paraleltermdd}
\end{align}   
and  
\begin{equation}    
 \widetilde{P}^T=-g^2T\sum_{n=-\infty}^{\infty}\sum_f\Big( \frac{\pi |q_fB|}{4\pi^2} \Big)e^{-\frac{q_\perp^2}{2|q_fB|}}\frac{q_\perp^2}{\vec{q}^2} \int \frac{dk_3}{(2\pi)} \left\{ 
\frac{-\tilde{\omega}_n^2+k_3^2}{[\tilde{\omega}_n^2+k_3^2+m_f^2][(\tilde{\omega}_n-\omega)^2+(k_3-q_3)^2+m_f^2]} \right\}. \label{paralelterm1dd}
\end{equation}    

We can therefore compute all the sums appearing in Eqs.~(\ref{paraleltermdd}) and~(\ref{paralelterm1dd}). Ignoring the vacuum contributions, we get
\begin{equation}
    \widetilde{P}^L=\frac{g^2}{64\pi^2}\sum_f |q_fB|e^{-\frac{q_\perp^2}{2|q_fB|}}\Bigg[ -2\omega^2+q^2-\frac{\omega^2q_3^2}{q^2} \Bigg] \Bigg( \frac{q_3^2}{\vec{q}^2q_\parallel^2}\Bigg)\Big(\frac{q_3}{T}\Big)
    \label{plhtl2d}
\end{equation}
and
\begin{equation}
    \widetilde{P}^T=-\frac{g^2}{64\pi^2}\sum_f |q_fB|e^{-\frac{q_\perp^2}{2|q_fB|}}\Bigg[ \frac{q_\perp^2 q_3^2}{q_\parallel^2 \vec{q}^2} \Bigg] \Big(\frac{q_3}{T}\Big).
    \label{pthtl2d}
\end{equation}

Finally, to obtain Eqs.~(\ref{plhtl211}) and~(\ref{pthtl2}), we perform the analytical continuation $i\omega\to q_0$, back to Minkowski space.

For consistency, we compute the coefficients $\widetilde{P}^L$ and $\widetilde{P}^T$
in an alternative way. First, in Euclidean space, we add and subtract the term $k_3^2+m_f^2$ in the numerators, such that the integrands look like
\begin{align}
    \widetilde{P}^L&=-g^2\sum_f\Big( \frac{\pi |q_fB|}{4\pi^2} \Big)e^{-\frac{q_\perp^2}{2|q_fB|}}\nonumber \\
    &\times \text{i} \ T\sum_n \int \frac{dk_3}{2\pi} \Bigg\{ \frac{(\tilde{\omega}_n^2+k_3^2+m_f^2)2\frac{(i\omega)^2}{\vec{q}^2}}{[\tilde{\omega}_n^2+k_3^2+m_f^2][(\tilde{\omega}_n-\omega)^2+(k_3-q_3)^2+m_f^2]}\nonumber \\
    &-\frac{4(k_3^2-m_f^2)\frac{(i\omega)^2}{\vec{q}^2}}{[\tilde{\omega}_n^2+k_3^2+m_f^2][(\tilde{\omega}_n-\omega)^2+(k_3-q_3)^2+m_f^2]}\nonumber \\
    &-\frac{(\tilde{\omega}_n^2+k_3^2+m_f^2)((i\omega)^2+q_3^2)\frac{(i\omega)^2}{\vec{q}^2q^2}}{[\tilde{\omega}_n^2+k_3^2+m_f^2][(\tilde{\omega}_n-\omega)^2+(k_3-q_3)^2+m_f^2]}\nonumber \\
    &+\frac{[k_3^2((i\omega)^2+q_3^2)+(i\omega)^2m_f^2]\frac{2(i\omega)^2}{\vec{q}^2q^2}}{[\tilde{\omega}_n^2+k_3^2+m_f^2][(\tilde{\omega}_n-\omega)^2+(k_3-q_3)^2+m_f^2]}\nonumber \\
    &-\frac{(\tilde{\omega}_n^2+k_3^2+m_f^2)\frac{q^2}{\vec{q}^2}}{[\tilde{\omega}_n^2+k_3^2+m_f^2][(\tilde{\omega}_n-\omega)^2+(k_3-q_3)^2+m_f^2]}\nonumber \\
    &+\frac{2(k_3^2+m_f^2)\frac{q^2}{\vec{q}^2}}{[\tilde{\omega}_n^2+k_3^2+m_f^2][(\tilde{\omega}_n-\omega)^2+(k_3-q_3)^2+m_f^2]}\Bigg\}
    \label{addsubsL}
\end{align}
and
\begin{align}
    \widetilde{P}^T&=-g^2\sum_f\Big( \frac{\pi |q_fB|}{4\pi^2} \Big)e^{-\frac{q_\perp^2}{2|q_fB|}}\nonumber \\
    &\times \text{i} \ T\sum_n \int \frac{dk_3}{2\pi} \Bigg\{ \frac{(\tilde{\omega}_n^2+k_3^2+m_f^2)\frac{q_\perp^2}{\vec{q}^2}}{[\tilde{\omega}_n^2+k_3^2+m_f^2][(\tilde{\omega}_n-\omega)^2+(k_3-q_3)^2+m_f^2]}\nonumber \\
    &+\frac{(2k_3^2-k_3q_3)\frac{q_\perp^2}{\vec{q}^2}}{[\tilde{\omega}_n^2+k_3^2+m_f^2][(\tilde{\omega}_n-\omega)^2+(k_3-q_3)^2+m_f^2]}\Bigg\}.
    \label{addsubsT}
\end{align}

We simplify Eqs.~(\ref{addsubsL}) and~(\ref{addsubsT}), and also ignore the terms propotional to $m_f^2$, thus we obtain
\begin{align}
    \widetilde{P}^L&=-g^2\sum_f\Big( \frac{\pi |q_fB|}{4\pi^2} \Big)e^{-\frac{q_\perp^2}{2|q_fB|}}\nonumber \\
    &\times \text{i} \ T\sum_n \int \frac{dk_3}{2\pi} \Bigg\{ \frac{2\frac{(i\omega)^2}{\vec{q}^2}}{[(\tilde{\omega}_n-\omega)^2+(k_3-q_3)^2+m_f^2]}\nonumber \\
    &-\frac{4k_3^2\frac{(i\omega)^2}{\vec{q}^2}}{[\tilde{\omega}_n^2+k_3^2+m_f^2][(\tilde{\omega}_n-\omega)^2+(k_3-q_3)^2+m_f^2]}\nonumber \\
    &-\frac{((i\omega)^2+q_3^2)\frac{(i\omega)^2}{\vec{q}^2q^2}}{[(\tilde{\omega}_n-\omega)^2+(k_3-q_3)^2+m_f^2]}\nonumber \\
    &+\frac{k_3^2((i\omega)^2+q_3^2)\frac{2(i\omega)^2}{\vec{q}^2q^2}}{[\tilde{\omega}_n^2+k_3^2+m_f^2][(\tilde{\omega}-\omega)^2+(k_3-q_3)^2+m_f^2]}\nonumber \\
    &-\frac{\frac{q^2}{\vec{q}^2}}{[(\tilde{\omega}_n-\omega)^2+(k_3-q_3)^2+m_f^2]}\nonumber \\
    &+\frac{2k_3^2\frac{q^2}{\vec{q}^2}}{[\tilde{\omega}_n^2+k_3^2+m_f^2][(\tilde{\omega}_n-\omega)^2+(k_3-q_3)^2+m_f^2]}\Bigg\}
    \label{addsubsL2}
\end{align}
and
\begin{align}
    \widetilde{P}^T&=-g^2\sum_f\Big( \frac{\pi |q_fB|}{4\pi^2} \Big)e^{-\frac{q_\perp^2}{2|q_fB|}}\nonumber \\
    &\times \text{i} \ T\sum_n \int \frac{dk_3}{2\pi} \Bigg\{ \frac{\frac{q_\perp^2}{\vec{q}^2}}{[(\tilde{\omega}_n-\omega)^2+(k_3-q_3)^2+m_f^2]}\nonumber \\
    &+\frac{(2k_3^2-k_3q_3)\frac{q_\perp^2}{\vec{q}^2}}{[\tilde{\omega}_n^2+k_3^2+m_f^2][(\tilde{\omega}_n-\omega)^2+(k_3-q_3)^2+m_f^2]}\Bigg\}.
    \label{addsubsT2}
\end{align}
From Eqs.~(\ref{addsubsL2}) and~(\ref{addsubsT2}), we identify two kinds of sums over the Matsubara frequencies which are equal to Eqs.~(\ref{suma1}) and (\ref{suma2}), where the notation does not change. Since, we work with the hierarchy of energy scales $m_f^2\ll q^2\ll T^2$, the matter terms from the sums can be expressed in the following way
\begin{equation}
T\sum_n \frac{1}{[(\tilde{\omega}_n-\omega)^2+(k_3-q_3)^2+m_f^2]}\approx -\frac{1}{\sqrt{(k_3-q_3)^2}} \tilde{f}\big((k_3-q_3)^2\big)
\label{sum1matter}
\end{equation}
and
\begin{align}
    T\sum_n &\frac{1}{[\tilde{\omega}_n^2+k_3^2+m_f^2][(\tilde{\omega}_n-\omega)^2+(k_3-q_3)^2+m_f^2]} \approx \frac{1}{4k_3\sqrt{(k_3-q_3)^2}} \nonumber \\
    &\Bigg \{ \big(\tilde{f}(k_3)+\tilde{f}(\sqrt{(k_3-q_3)^2})\big )\Big( \frac{1}{i\omega-k_3-\sqrt{(k_3-q_3)^2}}-\frac{1}{i\omega+k_3+\sqrt{(k_3-q_3)^2}}\Big) \nonumber \\
    &+\big( \tilde{f}(k_3)-\tilde{f}(\sqrt{(k_3-q_3)^2}) \big)\bigg(\frac{1}{i\omega+k_3-\sqrt{(k_3-q_3)^2}}-\frac{1}{i\omega-k_3+\sqrt{(k_3-q_3)^2}} \bigg)\Bigg. ,
    \label{sum2matter}
\end{align}
where in Eqs.~(\ref{sum1matter}) and~(\ref{sum2matter}), we have ignored the smallest energy scale, namely, the fermion mass. We now need to consider that the external momentum $q$ is softer than the temperature, therefore we neglect subdominant terms and perform a Taylor expansion in $\tilde{f}$ around $q=0$ to get
\begin{equation}
    T\sum_n \frac{1}{[(\tilde{\omega}_n-\omega)^2+(k_3-q_3)^2+m_f^2]} \approx -\frac{1}{k_3}\tilde{f}(k_3)
    \label{matter1-1}
\end{equation}
and
\begin{align}
     &T\sum_n \frac{1}{[\tilde{\omega}_n^2+k_3^2+m_f^2][(\tilde{\omega}_n-\omega)^2+(k_3-q_3)^2+m_f^2]} \approx -\frac{1}{4k_3^2} \nonumber \\
     &\times \Bigg \{\frac{2}{k_3}\tilde{f}(k_3)-\frac{2q_3}{(i\omega)^2-q_3^2}\Bigg[ \Big( \frac{q_3}{T} \Big)e^{k_3/T}\tilde{f}(k_3)^2\nonumber \\
     &+\Big( \frac{q_3^2}{2k_3T} \Big)e^{k_3/T}\bigg( \frac{k_3}{T}(e^{k_3/T}-1)+(e^{k_3/T}+1) \bigg) \tilde{f}(k_3)^3 \Bigg]\Bigg\}. 
     \label{matter2-1}
\end{align}
Substituting Eqs.~(\ref{matter1-1}) and~(\ref{matter2-1}) into Eqs.~(\ref{addsubsL2}) and~(\ref{addsubsT2}), the pure matter contribution to the coefficients becomes
\begin{align}
    \widetilde{P}^L&=-g^2\sum_f \Big( \frac{\pi |q_fB|}{4\pi^2} \Big)e^{-\frac{q_\perp^2}{2|q_fB|}}\nonumber \\
    &\int \frac{dk_3}{2\pi} \Bigg\{-\frac{2(i\omega)^2}{\vec{q}^2}\Big(\frac{1}{k_3}\Big)\tilde{f}(k_3)+\frac{2(i\omega)^2}{\vec{q}^2}\Big(\frac{1}{k_3}\Big)\tilde{f}(k_3)\nonumber \\
    &-\frac{(i\omega)^2}{\vec{q}^2}\Big[ \Big( \frac{q_3}{T} \Big)e^{k_3/T}\tilde{f}(k_3)^2\nonumber \\
     &+\Big( \frac{q_3^2}{2k_3T} \Big)e^{k_3/T}\bigg( \frac{k_3}{T}(e^{k_3/T}-1)+(e^{k_3/T}+1) \bigg) \tilde{f}(k_3)^3 \Big] \nonumber\\
     &+\frac{(i\omega)^2}{\vec{q}^2q^2}((i\omega)^2+q_3^2)\Big(\frac{1}{k_3}\Big)\tilde{f}(k_3)-\frac{(i\omega)^2}{\vec{q}^2q^2}((i\omega)^2+q_3^2)\Big(\frac{1}{k_3}\Big)\tilde{f}(k_3)\nonumber \\
     &+\frac{(i\omega)^2}{\vec{q}^2q^2}((i\omega)^2+q_3^2)\Big[ \Big( \frac{q_3}{T} \Big)e^{k_3/T}\tilde{f}(k_3)^2\nonumber \\
     &+\Big( \frac{q_3^2}{2k_3T} \Big)e^{k_3/T}\bigg( \frac{k_3}{T}(e^{k_3/T}-1)+(e^{k_3/T}+1) \bigg) \tilde{f}(k_3)^3 \Big]\nonumber \\
     &+\frac{q^2}{\vec{q}^2}\Big(\frac{1}{k_3}\Big)\tilde{f}(k_3)-\frac{q^2}{\vec{q}^2}\Big(\frac{1}{k_3}\Big)\tilde{f}(k_3)\nonumber \\
     &+\frac{q^2}{\vec{q}^2}\Big[ \Big( \frac{q_3}{T} \Big)e^{k_3/T}\tilde{f}(k_3)^2\nonumber \\
     &+\Big( \frac{q_3^2}{2k_3T} \Big)e^{k_3/T}\bigg( \frac{k_4}{T}(e^{k_3/T}-1)+(e^{k_3/T}+1) \bigg) \tilde{f}(k_3)^3 \Big]\Bigg\}\nonumber \\
     &=g^2\sum_f \Big( \frac{\pi |q_fB|}{4\pi^2} \Big)e^{-\frac{q_\perp^2}{2|q_fB|}}\Big( \frac{q_3^2}{\vec{q}^2q_\parallel^2} \Big)\bigg[(i\omega)^2\bigg(\frac{1+q_3^2}{q^2}+2\bigg)+q^2\bigg]\Big(\frac{q_3}{16\pi T}\Big)
     \label{finalresult2}
\end{align}
and
\begin{align}
    \widetilde{P}^T&=g^2\sum_f \Big( \frac{\pi |q_fB|}{4\pi^2} \Big)e^{-\frac{q_\perp^2}{2|q_fB|}}\nonumber \\
    &\int \frac{dk_3}{2\pi} \Bigg\{ \frac{q_\perp^2}{\vec{q}^2}\Big( \frac{1}{k_3}\Big)\tilde{f}(k_3)-\frac{q_\perp^2}{\vec{q}^2}\Big( \frac{1}{k_3}\Big)\tilde{f}(k_3)\nonumber \\
    &+\frac{q_3q_\perp^2}{\vec{q}^2((i\omega)^2-q_3^2)}\Big[ \Big( \frac{q_3}{T} \Big)e^{k_3/T}\tilde{f}(k_3)^2\nonumber \\
     &+\Big( \frac{q_3^2}{2k_3T} \Big)e^{k_3/T}\bigg( \frac{k_3}{T}(e^{k_3/T}-1)+(e^{k_3/T}+1) \bigg) \tilde{f}(k_3)^3 \Big]\Bigg \}\nonumber \\
     &=-g^2\sum_f \Big( \frac{\pi |q_fB|}{4\pi^2} \Big)e^{-\frac{q_\perp^2}{2|q_fB|}} \Big(\frac{q_\perp^2q_3^2}{q_\parallel^2\vec{q}^2} \Big)\Big(\frac{q_3}{16\pi T}\Big).
     \label{finalresult1}
\end{align}

After straightforward algebra, we realize that Eqs.~(\ref{finalresult2}) and~(\ref{finalresult1}) are equal to Eqs.~(\ref{plhtl2d}) and~(\ref{pthtl2d}), respectively. Hence, both ways to compute the matter contributions of $\widetilde{P}^T$ and $\widetilde{P}^L$ in the HTL approximation, with the hierarchy of energy scales given by $m_f^2\ll q^2\ll T^2$, provide the same result.
\end{widetext}

\end{document}